\documentclass[%
  aip,
 amsmath,amssymb,
reprint,%
]{revtex4-2}
\usepackage{graphicx}
\usepackage{dcolumn}
\usepackage{bm}
\usepackage{color}
\usepackage{float}

\usepackage[utf8]{inputenc}
\usepackage[T1]{fontenc}
\usepackage{mathptmx}
\usepackage{subfig}
\usepackage[inline]{enumitem}
\usepackage{caption}
\newcommand{\bogus}[1]{{}}

\begin{document}


\title{Parallel electrostatic collisionless shocks in hot-cold ablative mixing plasmas}

\author{Yanzeng Zhang and Xian-Zhu Tang}%
\affiliation{Theoretical Division, Los Alamos National Laboratory, Los Alamos, New Mexico 87545, USA}


\begin{abstract}
Hot-cold ablative mixing plasmas are ubiquitous in astrophysical and
laboratory systems, where a cold/dense plasma is roughly in pressure
balance with a hot/dilute plasma.  Examples include the plasma thermal
quench during major disruptions in tokamaks, interaction between a central hot-spot and
the solid liner in an inertial confinement fusion (ICF)
capsule, and the formation of large-scale structures in galaxy
clusters. In such systems, a parallel electrostatic collisionless
shock forms and plays a critical role in both the thermal collapse of
the hot plasma and the ablative mixing of cold ions. The formation and
dynamics of such shocks are investigated by employing one-dimensional
VPIC simulations and theoretical analyses, revealing key differences
from the well-studied collisionless shocks where an over-pressured,
high-density plasma expands into a rarefied background. Notably, the
shock formation has a weak dependence on the plasma pressure, provided
that the density ratio between the cold and hot plasmas is large. Instead,
the shock is primarily governed by the plasma temperatures on both
sides.  The collisionless electron thermal conduction flux in both
upstream and downstream regions follows the free-streaming limit
itself, but its spatial gradient exhibits convective scaling, ensuring
the same characteristic length scale of the electron temperature and
density evolution.
\end{abstract}

\maketitle
 
\section{Introduction}
 
\begin{figure}[hbt]
\centering
\includegraphics[width=0.4\textwidth]{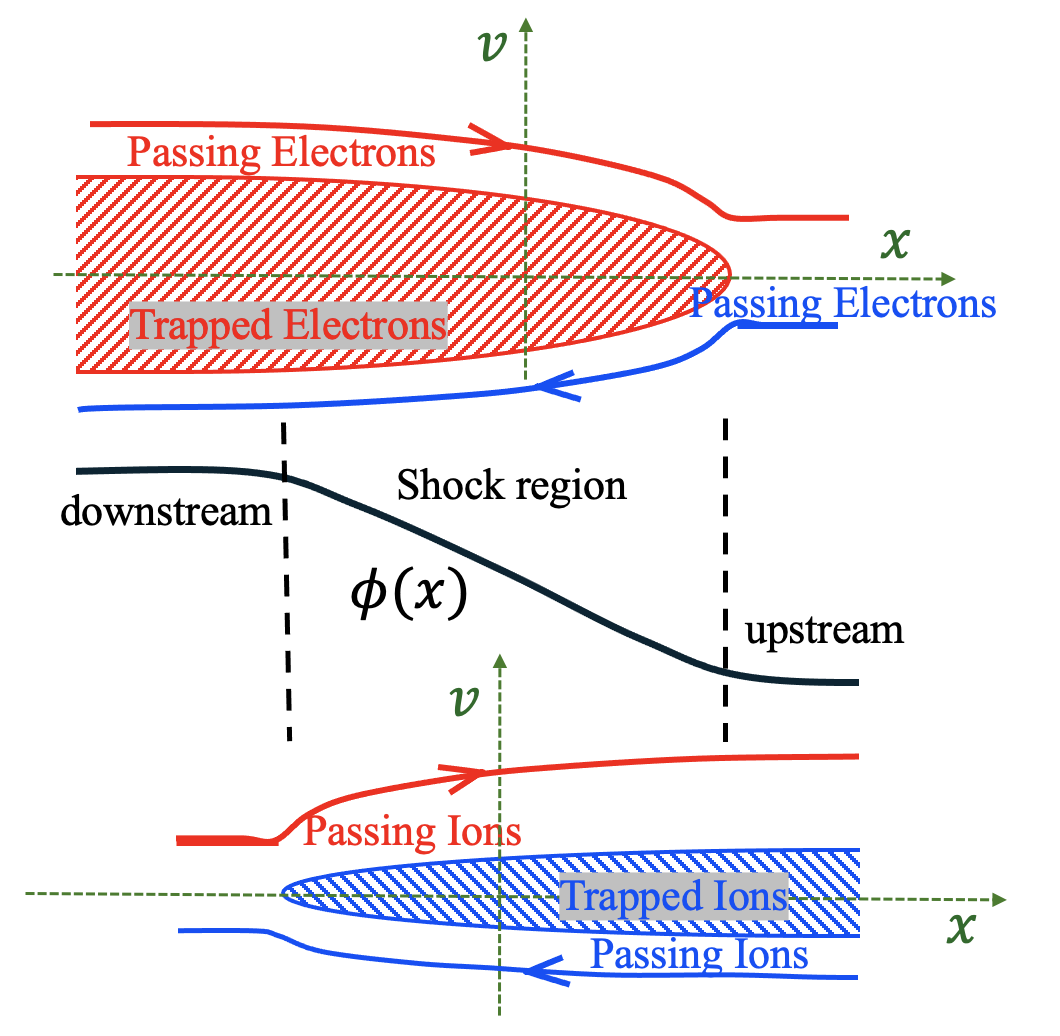}(a)
\includegraphics[width=0.4\textwidth]{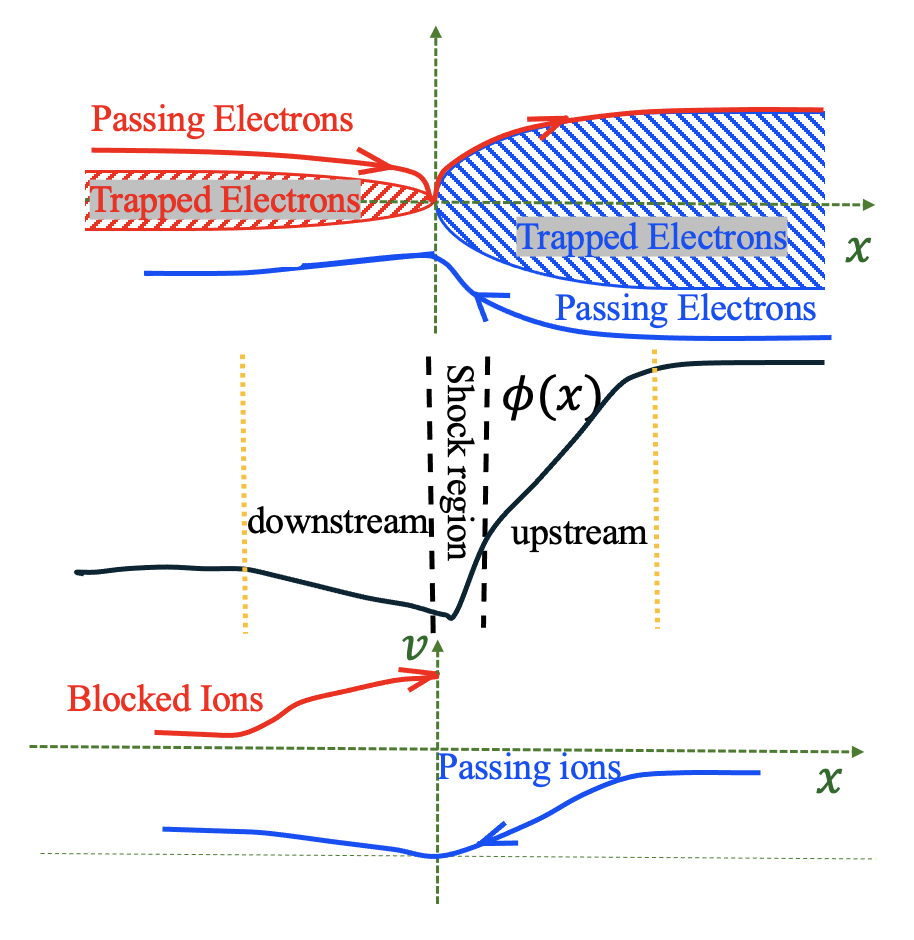}(b)
\caption{Schematic view of (a) classical shocks (upper panel) and (b)
  the ones in hot-cold ablative mixing plasmas (lower pannel). In each
  diagram, the electrostatic potential is sketched in the middle, and
  the electron and ion distribution functions are shown above and
  below the electrostatic potential sketch, respectively. The
  particles from downstream (upstream) are characterized by the red
  (blue) curves, where the trapped particles are shaded
  correspondingly. \textcolor{black}{The vertical dashed black lines characterize the shock region, while in (b) we also labeled the recession fronts by the vertical dotted yellow lines that primarily limit the electrostatic potential variation.}  }
\label{fig:diagram}
\end{figure}

Parallel collisionless shocks with respect to background magnetic
fields, or equivalently collisionless shocks in an unmagnetized
plasma, are prevalent in space, astrophysical, and laboratory plasmas
and play a crucial role in phenomena such as ion acceleration. Since
the 1960s, substantial efforts have been dedicated to this area
through experiments, theories, and simulations (e.g., see
Refs.~\onlinecite{sagdeev1966cooperative,moiseev1963collisionless,smith1970exact,forslund1970formation,andersen1967investigation,alikhanov1968zhetf,taylor1970observation,mason1971computer,montgomery1969shock,mozer1977observations,sorasio2006very,palmer2011monoenergetic,ji2008generating,giacalone1992hybrid},
and the review
papers~\onlinecite{tidman1971shock,biskamp1973collisionless,eselevich1982shock,ryutovl2018collisional,sakawa2016collisionless}
and references therein), and interest in this field has surged
recently due to the study of plasma interactions with high-power
lasers~\cite{koopman1967possible,haberberger2012collisionless,bell1988collisionless,wei2004ion,silva2004proton,romagnani2008observation,nilson2009generation,morita2010collisionless,fiuza2012laser,fiuza2013ion,zhang2017collisionless}.
Among these, the most well-studied case resembles an explosion in
which an over-pressured high-density plasma expands into the
surrounding rarefied plasma with a shock front, where the plasma
temperature, or more specifically the electron temperature, is
initially uniform or higher within the expanding
plasma~\cite{sarri2011generation,sarri2010shock,dieckmann2010simulation}
which is at the downstream of the shock, see the diagram in
Fig.~\ref{fig:diagram}(a). This leads to the formation of a
monotonically decreasing electrostatic potential from the downstream
to the upstream, capable of reflecting upstream ions for acceleration,
Fig.~\ref{fig:diagram}(a).
 
 A contrasting case, which is also ubiquitous in astrophysical and
 laboratory plasmas but less explored, has a cold/dense plasma roughly
 in pressure balance with a background hot/dilute plasma. This
 scenario occurs in, for example, the plasma thermal quench of a major
 disruption in
 tokamaks~\cite{zhang23cooling,zhang2023electron,li2023staged}, the
 interfacial dynamics between central hot-spot plasmas and the
 neighboring imploding liners or liner remnants in an inertial
 confinement fusion (ICF)
 capsule,~\cite{Lobatchev-Betti-PRL-2000,schiavi-atzeni-pop-2007,srinivasan-tang-pop-2014a,srinivasan-tang-epl-2014}
 and the formation of structures in galaxy
 clusters~\cite{Fabian-ARAA-1994,Peterson-Fabian-PR-2006}. In such
 cases, the hot-cold interface is known to trigger a thermal collapse
 of the nearly collisionless hot background plasma and an ablative mix
 of the cold ions with the background hot
 ions~\cite{zhang23cooling}. The latter becomes particularly
 interesting if they are of different species~\cite{ji2008generating},
 for example, in the high-Z impurity pellet injection into the fusion plasmas
 for tokamak disruption mitigation where efficient mixing of the
 ablated high-Z impurities with the fusion plasma is critical for
 sufficiently uniform radiations~\cite{mao2023rapid}.
 
 Recent kinetic studies of plasma thermal quench due to parallel
 transport have revealed the formation of a parallel collisionless
 shock in such hot-cold ablative mixing
 plasmas~\cite{zhang23cooling,zhang2023electron}. It is materialized
 by placing a nearly-collisionless fusion-grade plasma next to a
 cooling spot at the boundary that recycles plasma particles with a
 very low temperature, mimicking a radiative cooling spot that
 contains impurities.  Due to their pileup near the cooling spot,
 these cold recycled particles are nearly in pressure balance with the
 ambient hot plasma, forming a shock at which the cold recycled ions
 meet the background hot ions. Such shock exhibits a fundamentally
 different physical picture compared to the conventional collisionless
 shocks~\cite{sarri2011generation,sarri2010shock,dieckmann2010simulation}.
 A key distinction lies in the formation of a large, increasing
 electrostatic potential towards the upstream hot plasma, driven by
 the fast tail loss of the hot electrons, as illustrated in the
 diagram of Fig.~\ref{fig:diagram}(b). Rather than reflecting the
 upstream ions as in classic shocks, this potential accelerates them
 into the shock to be collisionlessly mixed with the downstream cold
 and dense ions.  In contrast, upstream hot electrons are trapped by
 the electrostatic potential, which significantly modifies the
 electron thermal conduction heat flux and hence the plasma cooling
 process~\cite{zhang23cooling}.

In this paper, we will study in detail the generic properties of this
distinct class of shocks, focusing on their structures and speed, and
the plasma dynamics in both upstream and downstream regions next to
the shock front. To achieve this, one-dimensional (1D)
first-principles kinetic simulations using the VPIC code~\cite{VPIC}
will be employed. Unlike previous thermal quench studies that
incorporated a cooling boundary
condition~\cite{zhang23cooling,zhang2023electron} for obtaining the
cold plasma, we directly initialize a cold, dense plasma interacting
with a hot, dilute plasma following the standard shock
setup~\cite{sarri2010shock}. This effectively removes the boundary
condition effects. A striking finding is that the formation of the
shock has a weak dependence on the initial plasma pressure, provided
that the density ratio is large, $n_c/n_h\gg \sqrt{T_h/T_c}$, where
$n_{c,h}$ and $T_{c,h}$ are the initial densities and temperatures of
the cold and hot plasmas, respectively. Instead, the shock front speed
is primarily governed by $T_c$, while the shock width is determined by
$T_h$.  The expansion of the cold plasma is sustained by a recession
front further downstream, analogous to the recession front in the
upstream hot plasma~\cite{zhang23cooling}, as illustrated in
Fig.~\ref{fig:diagram}(b). These two recession fronts
independently drive the density collapse in the cold and hot plasmas,
leading to the formation of a non-monotonic electrostatic potential
between them. Interestingly, the collisionless electron
thermal conduction flux in both upstream and downstream regions
adheres to the free-streaming limit that scales with the electron
thermal speed, yet its spatial gradient follows a convective scaling
that is proportional to ion flow.  The convective scaling of electron
thermal conduction flux ensures that the electron temperature and
density evolve over comparable spatial scales.

The rest of the paper is organized as follows. Section \ref{sec:setup}
details the VPIC simulation setup. Section~\ref{sec:shock-structure}
presents the physical picture of the shock, discussing its formation,
shock front width, shock speed, and the behavior of plasma thermal
conduction heat fluxes. In Section~\ref{sec:model}, we provide a
theoretical analysis to interpret and support the simulation results.
Section \ref{sec:conclusion} will conclude.

\section{VPIC simulation setup\label{sec:setup}}

We use fully kinetic simulations with the VPIC code~\cite{VPIC} to
investigate the properties of the parallel collisionless shocks in
hot-cold ablative mixing plasmas. A 1D slab geometry is considered,
where hot and cold plasmas are initialized at $x>0$ and $x<0$,
respectively. Unless otherwise specified, both plasmas consist of
equilibrium electrons and hydrogen ions, i.e.,
$n_e^{h,c}=n_i^{h,c}=n_{h,c}$ and $T_e^{h,c}=T_i^{h,c}=T_{h,c}$ with
$n_h$ and $T_{h}$ ($n_c$ and $T_c$) the initial density and
temperature of the hot (cold) plasma, respectively. In addition, the
system is initialized in pressure balance, $p_0=n_hT_h=n_cT_c$, with a
large temperature and density ratio $R=T_{h}/T_c = n_{c}/n_h \gg 1$
(in this study, we consider $R\in[10, 10^3]$).  We shall emphasize
that, while the initial pressure balance simplifies the setup, it is
the temperature ratio that plays a critical role in the shock
formation as will be discussed in the next section.

The boundaries of the simulation domain are thermobaths that recycle
all the particles drawn from half-Maxwellian distributions with
temperatures $T_h$ and $T_c$ at the right and left boundaries,
respectively. These boundaries are placed far from $x=0$ to minimize
their influences on the shock properties. That is to say, we
effectively simulate an infinite plasma.

In such a setup, the only intrinsic length scale of the system is the
Debye length, $\lambda_D\propto \sqrt{T_e/n_e}$, which is $R$ times
larger in the hot plasma than that in the cold plasma,
$\lambda_D^h=R\lambda_D^c$. Correspondingly, the plasma frequency in
the hot plasma is reduced by $\sqrt{R}$,
$\omega_{pe}^h=\omega_{pe}^c/\sqrt{R}$. Therefore, the VPIC
simulations need to resolve the cold plasma Debye length $\lambda_D^c$
and plasma frequency $\omega_{pe}^c$. We have chosen a grid size of
$\Delta x= \lambda_D^c$, while the time
step $\Delta t$ is determined accordingly. Whereas, the minimum number
of particles per cell is limited by the hot plasma if the same
particle weight is considered, which will cause significant
computational costs considering the large density ratio. To address
this, we employ a weighted particle scheme, assigning the cold plasma
particles a weight $R$ times that of the hot plasma particles,
ensuring an initial uniform particle count of $10^4$ particles per
cell throughout the domain.

We performed both electrostatic and electromagnetic simulations, with
or without a uniform guide magnetic field $B_x$ (with electron
gyrofrequency equal to the hot plasma frequency
$\omega_{ce}=\omega_{pe}^h$). The results show that the shock
formation is unaffected, revealing that such shock is electrostatic in
nature. Therefore, the following analysis focuses on the electrostatic
simulations without an external magnetic field.
The physics analysis follows the standard definition of
plasma density, parallel flow, parallel temperature
and thermal conduction heat flux of the parallel degrees
in statistical physics, which 
can be
calculated from the particle distribution function $f$ as
\begin{align}
 n=&\int f d^3v, \\
nV_x=&\int v_x f d^3v, \\
 nT_x=& m\int (v_x-V_x)^2f d^3v, \\
 q_{n}=&m\int (v_x-V_x)^3 f d^3v.
\end{align}

\section{The physical picture of the shock \label{sec:shock-structure}}
\subsection{The formation of the shock}

\begin{figure}[hbt]
\centering
\includegraphics[width=0.45\textwidth]{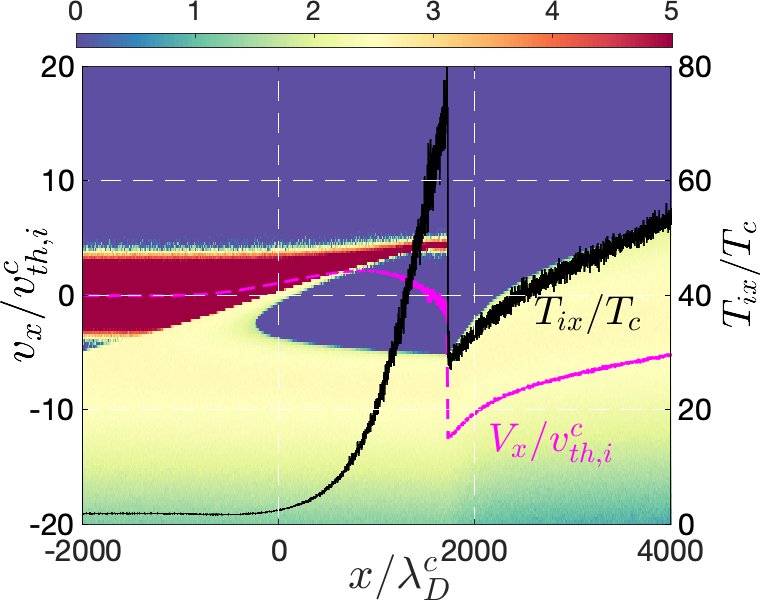}
\includegraphics[width=0.45\textwidth]{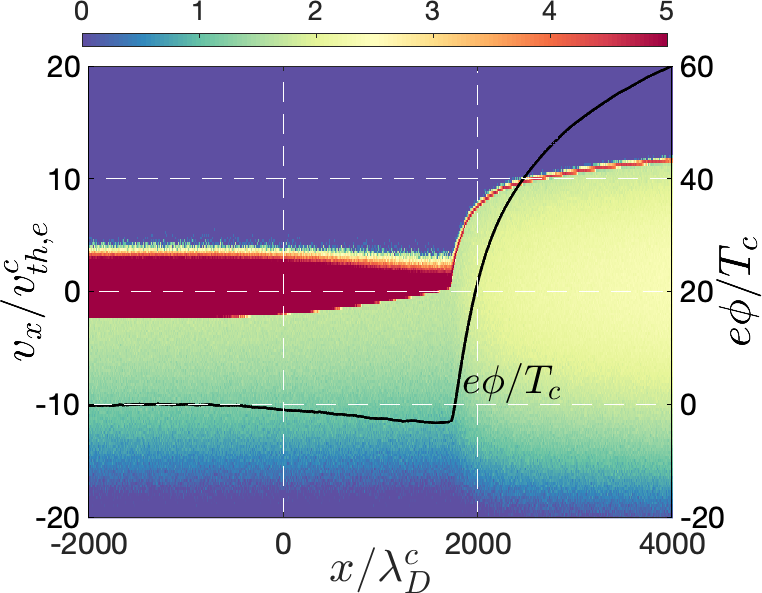}
\caption{Phasespace of ions (upper panel) and electrons (lower panel)
  at $\omega_{pe}^ct\approx 4\times 10^3$ for $R=100$, where the
  colormap (with a cutoff at 5) shows the logarithm of the particle
  distribution. In the ion phasespace plot, we also show the parallel
  ion flow $V_{ix}$ corresponding to the left y-axis and the parallel
  ion temperature $T_{ix}$ to the right y-axis, while in the electron
  plot, we also show the electrostatic potential $\phi$ corresponding
  to the right y-axis. Here all the normalized quantities are from the
  cod plasma, with
  $v_{th,e,i}^c\equiv\sqrt{T_{e,i}^c/m_{e,i}}=\sqrt{T_c/m_{e,i}}$ the
  cold electron and ion thermal speeds. Reduced ion mass of
  $m_i/m_e=100$ is used in the simulation.}
\label{fig:fvx-x}
\end{figure}

The initial hot-cold plasma interface triggers a thermal collapse of
the hot plasma. This collapse is primarily due to the loss of
high-energy or tail electrons from the hot plasma to the cold region,
which generates an electric field that traps hot electrons by
reflecting  their thermal bulk to maintain the ambipolarity (see the
\textcolor{black}{diagram} in Fig.~\ref{fig:diagram}(b) and the electron
phase space in Fig.~\ref{fig:fvx-x}). This ambipolar electric field is
bounded by an upstream recession front that governs the collapse of both the
plasma density and the ion parallel temperature as in the thermal
quench case~\cite{zhang23cooling}.  Since the upstream plasma remains
unaffected by the cold ions, the upstream recession front speed should match
that found in Ref.~\onlinecite{zhang23cooling}, given by
$U_{RF}^{up}\approx 3v_{th,i}^h$ with
$v_{th,i}^h\equiv\sqrt{T_i^h/m_i}=\sqrt{T_h/m_i}$ the thermal speed of
the hot ions. The potential drop between this recession front at
$x^{rh}$ and the hot-cold interface, defined as the cold ion front for
$t>0$ (located at $x^{sh}\approx 1730\lambda_D^c$ in
Fig.~\ref{fig:fvx-x}), is primarily set by the hot electron
temperature, following $e\Delta \phi^{up} \equiv e\left(\Phi(x^{rh}) -
\Phi(x^{sh})\right)\sim T_h$ as illustrated in
Fig.~\ref{fig:fvx-x}. This potential accelerates the hot ions toward
the interface at $x^{sh},$ generating a plasma flow with a speed on
the order of the hot ion sound speed, $V_{ix}^{up}\sim c_s^h\sim
v_{th,i}^h$.

\begin{figure}[hbt]
\centering
\includegraphics[width=0.45\textwidth]{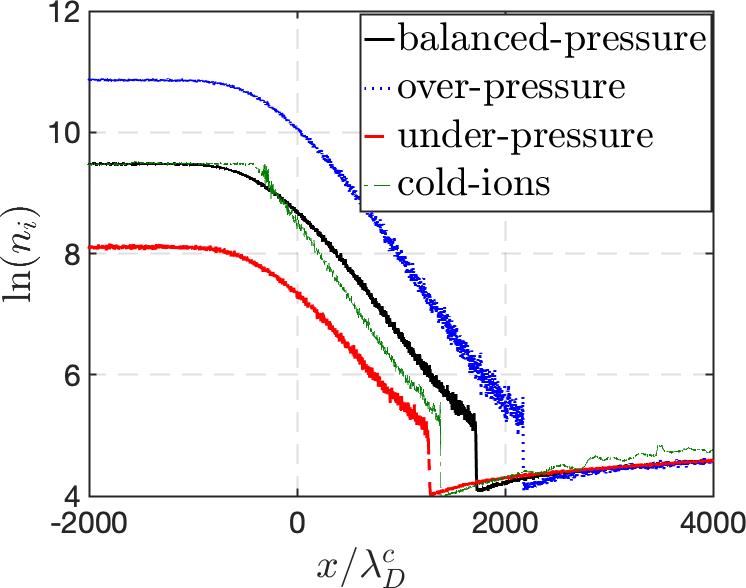}
\caption{Total ion density at the same time for different initial cold
  plasma pressure adjusted by the plasma density and the ``cold'' ion
  simulation. The balanced-pressure is the same as
  Fig.~\ref{fig:fvx-x}, while the cold plasma density in the
  over-pressure (under-pressure) case is enlarged (reduced) by a
  factor of 4. In the ``cold'' ion simulation, both the upstream and
  downstream ions are initially replaced by extremely cold ions with temperature of
  $T_c/100$. Here $\lambda_D^c$ is the cold plasma Debye length in
  the balanced-pressure case.}
\label{fig:ne-diff-pressure}
\end{figure}

Similar to the upstream hot plasma, the expansion of the cold plasma
toward the upstream is sustained by a recession front in the
downstream located at $x^{rc}\approx-1200\lambda_D^c$ in
Fig.~\ref{fig:fvx-x} (also the solid black line in
Fig.~\ref{fig:ne-diff-pressure}), which governs the collapse of the
cold plasma density. Due to the density drop, a decreasing
electrostatic potential forms between this cold or downstream recession front and
the hot-cold interface, leading to the trapping of cold electrons and
the acceleration of cold ions in the downstream region toward the hot-cold interface. This results
in an electrostatic potential structure between the two recession
fronts that resembles a dynamic, non-monotonic double
layer~\cite{sato1981numerical,jm1995dynamics,kato2010electrostatic}. In
the downstream, the electron heating, driven mainly by the mixing of
cold and hot components, partially offsets the electron density drop,
leading to a modest potential drop set by the cold plasma temperature,
$- e\Delta \phi^{dn} = - e\left(\Phi(x^{sh}) - \Phi(x^{rc}\right) \sim
T_c\ll T_h$. Consequently, hot ions and electrons in the downstream
experience minimal changes in velocity as they move away from the
hot-cold interface at $x^{sh}:$ the hot ions undergo slight
deceleration, while hot electrons are mildly accelerated, with
velocity changes scaling as $|\Delta v_{i,e}|\sim \sqrt{- e\Delta
  \phi^{dn}/m_{i,e}}\ll v_{th,i,e}^h$ (as shown in the colormap of
their phase space plots in Fig.~\ref{fig:fvx-x}).

Fig.~\ref{fig:fvx-x} shows that a shock front forms at the hot-cold
interface, marked by the cold plasma tip (the cold plasma appears as
the red color in the phase space due to its high density). Across the
shock front, all key plasma state variables (plasma density,
flow, and temperature) exhibit distinct jumps. Particularly,
substantial ion flow energy from the upstream is transformed into ion
thermal energy through the collisionless mixing of hot and cold ions
downstream right behind the shock. Here, the flow of the cold ions, accelerated by the
potential drop $\Delta \phi^{dn}$, nearly balances the plasma flow
driven by the incoming hot ions so the total plasma flow
approximately vanishes.  It is worth noting that, during the
acceleration, the cold ions experience decompressional cooling,
resulting in the formation of a cold ion beam near the shock front.
 
In contrast to the cold ions that are completely blocked by the shock
front and thus confined to the downstream region, some cold electrons
manage to penetrate through the shock front and are then accelerated by the upstream
electrostatic potential. Notably, this is the same electrostatic
potential responsible for trapping the hot electrons. Thus, the cold
electrons form a beam at the trap-passing boundary in the hot electron
distribution. This can be seen in the thin red stripe in the upstream phase space
plot for electrons in Fig.~\ref{fig:fvx-x}.

The physical picture given above indicates that the electrostatic
potential plays a critical role in the formation of this class of shocks.
Since ambipolar potential is
primarily set by the effective electron pressure, we find two distinct characteristic
properties of the shock. First, since the hot electrons can
move nearly freely into the cold plasma with $-e\Delta \phi^{dn}\ll
T_h$, the shock formation should only have weak dependence on the
initial cold plasma pressure. That is, the shock persists across a
wide range of downstream conditions--including balanced-pressure,
over-pressure, and even under-pressure cases relative to the upstream
hot plasma--as long as the temperature ratio remains high. This has
been confirmed by simulations with varying downstream cold plasma
densities while maintaining a fixed temperature ratio of
$R=T_h/T_c=100$ as shown in Fig.~\ref{fig:ne-diff-pressure}. In
particular, it demonstrates that both the structure and speed of the
shock front are only weakly influenced by the cold plasma density.

Second, the ion pressure or more specifically the ion
temperature in both upstream and downstream should have minimal impact
on the shock formation. This has been confirmed by the ``cold'' ion
simulation in Fig.~\ref{fig:ne-diff-pressure}, where both upstream and
downstream ions are initialized with an extremely low temperature of
$T_i=T_c/100$. This independence on the ion temperature is consistent
with the expanding shock case, where uniform cold ions are initialized
as well~\cite{sarri2011generation}. As a result, the characteristic
velocities for ions are actually the ion sound speeds
($\propto\sqrt{T_e/m_i}$) rather than the ion thermal speeds
($\propto\sqrt{T_i/m_i}$), although they are comparable for the
non-cold-ion cases since $T_i=T_e$ initially in both downstream and
upstream regions. The main difference in the cold-ion simulation is that the
hot ions behind the upstream recession front form a beam as well,
which can be unstable~\cite{buneman1958instability} as indicated by
the density perturbations in Fig.~\ref{fig:ne-diff-pressure}.

\subsection{Shock front width and speed}

\begin{figure}[hbt]
\centering
\includegraphics[width=0.45\textwidth]{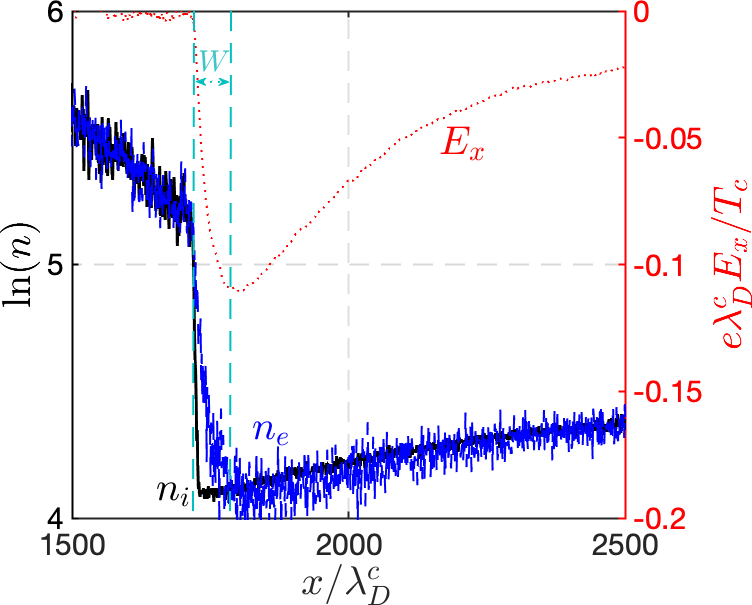}
\caption{Ion and electron density (left y-axis) as well as the
  electric field (right y-axis) corresponding to
  Fig.~\ref{fig:fvx-x}. Two vertical cyan lines separate the
  quasi-neutral plasmas (the ion layers) away from and the non-neutral
  plasma (the electron layer) at the shock front. The width of the
  latter is at the order of hot plasma Debye length,
  $W\sim\lambda_D^h=100\lambda_D^c.$ }
\label{fig:ni-ne-Ex}
\end{figure}

The non-monotonic electrostatic potential profile reveals the presence
of three distinct charge layers: two ion layers located away from the
shock front and a central electron layer near the shock front (e.g.,
see Fig.~\ref{fig:ni-ne-Ex}). The ion layers are limited by the two
recession fronts excluding a narrow central electron layer around the
shock front.  In these two ion layers, quasineutrality is weakly
violated, just like the transition layer that connects the
quasineutral presheath to the non-neutral sheath
plasmas~\cite{Li-etal-prl-2022,li2022transport}. In contrast, the
electron layer near the shock front is strongly non-neutral,
characterized by an electron density that significantly exceeds the
ion density. The width of this layer is marked by ``W'' in
Fig.~\ref{fig:ni-ne-Ex}.  As a result, the electric field undergoes a
sharp transition within this layer, decreasing from its upstream value
of $E_x^{up}\sim T_c/(e\lambda_D^c)=T_h/(e\lambda_D^h)$ to nearly zero
downstream, as illustrated in Fig.~\ref{fig:ni-ne-Ex}.

\begin{figure}[hbt]
\centering
\includegraphics[width=0.4\textwidth]{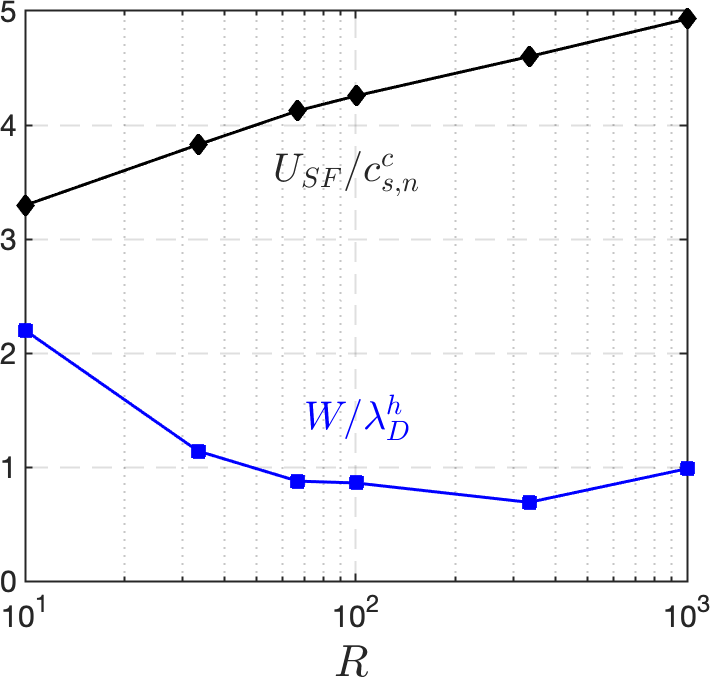}
\caption{The shock front width $W$ and speed $U_{SF}$ for different $R=T_h/T_c=n_c/n_h$. The rest simulation setup is the same as Fig.~\ref{fig:fvx-x}. $W$ is obtained using the method in
  Fig.~\ref{fig:ni-ne-Ex} and $U_{SF}$ is normalized by the nominal cold ion sound speed $c_{s,n}^c.$}
\label{fig:cooling-front-width}
\end{figure}

The width of the non-neutral (electron) layer is on the order of the
hot plasma Debye length due to the Debye
shielding~\cite{chen2012introduction}. This characteristic width
effectively defines the shock front thickness, leading to the
conclusion that the shock front width scales with the upstream plasma
Debye length, consistent with conventional parallel collisionless
shocks~\cite{moiseev1963collisionless,smith1970exact,taylor1970observation,romagnani2008observation,forslund1970formation,sarri2010shock}. This
relationship has been verified by simulations across various
temperature and density ratios ($R$) in the balanced-pressure setup,
as shown in Fig.~\ref{fig:cooling-front-width}.

In contrast, the shock front speed is primarily governed by the cold
plasma temperature, or more specifically the cold
  electron temperature. This can be seen in the cold-ion case in
  Fig.~\ref{fig:ne-diff-pressure}, where the shock front speed has a weak dependency on the cold
plasma density (e.g., see
Fig.~\ref{fig:ne-diff-pressure}). Therefore, it's
  natural to normalize the shock front speed using the cold ion sound
  speed and the nominal one is 
  \begin{equation}
  c_{s,n}^c\equiv \sqrt{T_c/m_i}.\label{eq-cold-ion-sound-nominal}
  \end{equation} In
  the cases where the initial temperatures of cold electrons and ions
  are the same, one has $c_{s,n}^c=v_{th,i}^c$. Otherwise, in the
  cold-ion case with $T_i\ll T_c$, the electron temperature dominate
  the cold ion sound speed, $c_{s,n}^c\gg v_{th,i}^c$. The simulations
  in Fig.~\ref{fig:cooling-front-width} show that $U_{SF}\sim 4
  c_{s,n}^c$ for the former case. This scaling aligns
with the observation that the electrostatic potential drop in the
downstream region satisfies $- e\Delta \phi^{dn}\sim T_c.$
\textcolor{black}{It is important to note that the shock front speed
  scaling is different from predictions based on discontinuity
  analysis that neglects plasma conduction heat
  fluxes~\cite{zhang23cooling,zhang2023electron}, which suggest a
  front speed on the order of the upstream plasma sound speed. This
  discrepancy highlights the critical role of collisionless conduction
  heat flux in accurately describing the shock dynamics. As will be
  discussed in the next section, the heat flux exhibits significant
  discontinuities across the shock front, substantially affecting its
  structure and propagation speed. We also note that the inclusion of
  collisions in the thermal quench can alter the shock behavior, which
  tends to smooth the shock front and accelerate its propagation.
}

\subsection{Thermal conduction heat fluxes}

The electron thermal conduction heat flux, $q_{en}\equiv m_e\int
(v_x-V_{ex})^3f_e(v_x)dv_x$, plays a crucial role in the evolution of
the parallel electron temperature ($T_{ex}$). Recent studies on the
plasma thermal quench~\cite{zhang23cooling} reveal that $q_{en}$ is
significantly influenced by ambipolar transport constraint between the
upstream recession front and the shock front. As such, $q_{en}$ itself
follows the free-streaming limit~\cite{atzeni_book_2004,Bell-pof-1985},
\begin{align} 
q_{en} \approx \alpha n_h  T_h v_{th,e}^h\equiv\alpha p_0v_{th,e}^h, \label{eq:qe-flux-limiter} 
\end{align}
with $\alpha \sim 0.1$, but its spatial gradient exhibits the
convective scaling~\cite{zhang23cooling}
\begin{align}
  dq_{en}/dx\propto V_{ex} \approx V_{ix} \propto 1/\sqrt{m_i},
\end{align}
with $V_{ex}$ and $V_{ix}$ the electron and ion flow, respectively.

\begin{figure}[hbt]
\centering
\includegraphics[width=0.45\textwidth]{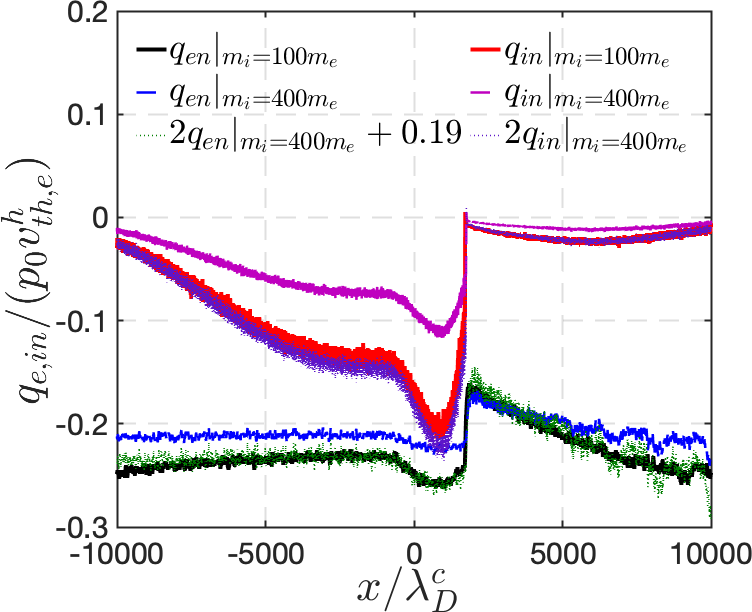}
\caption{Normalized electron ($q_{en}$) and ion ($q_{in}$) conduction
  heat flux for $m_i=100m_e$ and $m_i=400m_e$. For the latter, we also
  plot $2q_{en}+0.19$ and $2q_{in}$. The rest parameters are the same
  as Fig.~\ref{fig:fvx-x}.}
\label{fig:condu-diff-mass}
\end{figure}

What we find here is that this behavior of $q_{en}$ holds throughout
the entire plasma region between the downstream and upstream recession
fronts, including near the shock front. This has been demonstrated in
Fig.~\ref{fig:condu-diff-mass} by comparing the black curve with the blue
one for $q_{en}$ and the black curve with the green one for
$dq_{en}/dx.$ As a result, the electron temperature profile evolves
over the same characteristic length scale as the electron density in
both the upstream and downstream regions. In contrast, the ion
conduction heat flux itself follows the convective scaling
\begin{align}
  q_{in}\propto V_x\propto 1/\sqrt{m_i},
\end{align}
as shown in Fig.~\ref{fig:condu-diff-mass}.

\section{Theoretical analysis \label{sec:model}}

In this section, we provide a theoretical analysis of the findings in
the previous section. For convenience, we will use the superscript
$up$ ($dn$) to denote the upstream (downstream) plasma moments,
including the density, flow, temperature and conduction heat flux. The
quantities related to the hot and cold particle component are denoted
by $h$ and $c$, respectively. Moreover, we will use $A^{up,dn}(\rm
SF)$ to denote the quantity $A$ near the entrance, from upstream or
downstream, respectively, to the non-neutral layer in Fig.~\ref{fig:ni-ne-Ex}.

\subsection{Electron dynamics}
The upstream electron dynamics, away from the non-neutral layer in Fig.~\ref{fig:ni-ne-Ex}, is the same as the thermal quench case~\cite{zhang2023electron}, for which one can assume the distribution function as the sum of a truncated Maxwellian (for the trapped hot electrons) and cold electron beam. 
Considered 1D case, it reads
\begin{align}
f_{e}^{up} &= f_{et}^{h} + f_{eb}^{c}, \label{eq:up-fe}\\
f_{et}^{h}&= \frac{n_{et}^{h}}{\sqrt{2\pi}\alpha_{e}^h
  v_{th,e}^h} e^{-v_x^2/2(v_{th,e}^h)^2}
\Theta\left(1-\frac{v_x}{v_{ec}^{h}}\right),\label{eq:up-fe-trap}\\
f_{eb}^{up}&= n_{eb}^c\delta(v_x -
v_{ec}^{h}),\label{eq:up-fe-beam}
\end{align} 
where $v_{ec}^{h}=\sqrt{2e\Delta \phi/m_e}$ with $\Delta
\phi=\phi(x)-\phi_{\rm min}$ and $\phi_{\rm min}$ being the minimum
potential at the shock front,
$\alpha_{e}^h=[1+\textup{Erf}(v_{ec}^{h}/\sqrt{2}v_{th,e}^h)]/2$,
$\Theta(x)$ is the Heaviside step function that vanishes for $x<0$,
$\delta(x)$ the Dirac delta function, $\textup{Erf}(x)$ is the error
function, and $n_{et}^{h}$ and $n_{eb}^c$ denote the density of the
trapped hot and cold beam electrons, respectively.

The plasma density drop is limited by the electrostatic potential, where
\begin{equation}
n_{e}^{up}=n_{et}^{h}+n_{eb}^c\sim n_h.\label{eq:ne-coldbeam}
\end{equation}
 The constraint of the ambipolarity requires that the electron flow is approximately the same as the ion's, i.e., $V_{ex}^{up}\approx V_{ix}^{up}\sim c_s^h\sim \sqrt{T_h/m_i}$, where
\begin{align}
 n_{e}^{up}V_{ex}^{up}=-\frac{n_{et}^{h}v_{th,e}^h}{\sqrt{2\pi}\alpha_e^{h}} e^{-(v_{ec}^{h})^2/2(v_{th,e}^h)^2}+n_{eb}^cv_{ec}^{h}. \label{eq:nV-cold-beam}
\end{align}
In the large cold beam component limit $n_{eb}^c /n_{e}^{up}\gg \sqrt{m_e/m_i}$, one has
\begin{align}
n_{eb}^cv_{ec}^{h}\approx \frac{n_{et}^{h}v_{th,e}^h}{\sqrt{2\pi}\alpha_e^{h}} e^{-(v_{ec}^{h})^2/2(v_{th,e}^h)^2}, \label{eq:cold-e-beam-flux}
\end{align}
away from the shock front since $v_{ec}^{h}\sim v_{th,e}^h\gg V_{ex}^{up}$. It follows that, approaching the shock front, $n_{et}^{h}/n_{eb}^c$ is decreasing with $v_{ec}^{h}$. However, $v_{ec}^{h}$ has a minimal limit that is set by the electrostatic potential in the non-neutral layer in Fig.~\ref{fig:ni-ne-Ex}. To obtain its scaling, we assume that $E_x$ is linearly increased from zero to $eE_x\sim T_h/\lambda_D^h$ in such layer with $W\sim \lambda_D^h$, resulting in the potential increase in such layer of $\Delta \phi_{nl}\approx T_h/2$. As a result, the minimal value of $v_{ec}^{h}$ is set as $v_{ec}^{h}\sim \sqrt{2e\Delta \phi_{nl}/m_e}\sim v_{th,e}^h$ so that $n_{et}^{h}\sim n_{eb}^c\sim n_h$ near the shock front. Simulations that separate the hot and cold electrons as different species have confirmed that $(v_{ec}^{h})^{up}({\rm SF})\approx 0.36v_{th,e}^h$ and $(n_{et}^{h})^{up}({\rm SF})\approx 0.26 n_h$ for Fig.~\ref{fig:fvx-x}.

The electron temperature reads
\begin{equation}
  T_{ex}^{up} = T_h \left[
    1 - \frac{n_{eb}^c}{n_{e}^{up}} + \frac{V_{ex}^{up}v_{ec}^{h}}{(v_{th,e}^h)^2} - \frac{(V_{ex}^{up})^2}{(v_{th,e}^h)^2}
    \right], \label{eq:Tepara-up}
\end{equation}
so that $T_{ex}^{up} ({\rm SF})\sim T_h$. Specifically, for
Fig.~\ref{fig:fvx-x}, we have $T_{ex}^{up}\sim 0.5T_h$ with
$n_e^{up}({\rm SF})\sim 0.5 n_h$ and $(n_{et}^{h})^{up}({\rm
  SF})\approx 0.26 n_h$.

The thermal conduction heat flux reads~\cite{zhang2023electron}
\begin{align}
   q_{en}^{up}&\approx -2n_{eb}^cv_{ec}^{h}T_h+ \left[\left(\frac{(v_{ec}^{h})^2}{(v_{th,e}^h)^2}+2\right)\frac{T_h}{T_{ex}^{up}} -3\right]
  n_e^{up}V_{ex}^{up}T_{ex}^{up},\label{eq:qen-up}
\end{align}
where we have dropped the small term of $- n_e^{up} m_e
(V_{ex}^{up})^3$. The first term in $q_{en}^{up}$, following the
free-streaming limit, dominates over the second term that has the
convective scaling since $n_{eb}^cv_{ec}^{h}\sim n_hv_{th,e}^h\gg
n_{e}^{up}V_{ex}^{up}\sim n_hc_{s}^h$. However, the spatial
gradient of $n_{eb}^cv_{ec}^{h}$ is set by $n_e^{up}V_{ex}^{up}$ as
indicated by Eq.~(\ref{eq:nV-cold-beam}) so that the
$dq_{en}^{up}/dx\propto V_{ex}^{up} \approx V_{ix}^{up} \propto 1/\sqrt{m_i}$. Near the
shock front, we have $q_{en}^{up}({\rm SF})\approx -2\times 0.25n_h
\times 0.36 v_{th,e}^h\times T_h\approx -0.18p_0v_{th,e}^h$, in agreement
with the simulation data in Fig.~\ref{fig:condu-diff-mass}.

In the non-neutral electron layer, $n_{et}^{h} $ is further reduced to
$(n_{et}^{h})^{dn}({\rm SF})\approx 0.2n_h$ due to the potential drop
$\Delta\phi_{nl}$ and so are $n_{et}^{h}/n_{eb}^c$ and $v_{ec}^{h}$.

In the downstream outside the non-neutral layer, the variations of
$v_{ec}^{h}$ and $n_{et}^{h}$ is small since $e\Delta\phi^{dn}\sim
T_c\ll T_h$. As a result, the hot electron distribution is
approximately a half-Maxwellian with a nearly constant density, i.e.,
$f_{et}^{h}$ in Eq.~(\ref{eq:up-fe-trap}) with
$v_{ec}^{h}/v_{th,e}^h\approx 0$ and $n_{et}^{h}\approx 0.2 n_h$. On
the other hand, the cold electron distribution can be assumed as a
cutoff-Maxwellian as well (e.g., see Fig.~\ref{fig:fvx-x})
\begin{equation}
f_{et}^{c}=\frac{n_{et}^{c}\left(\Delta \phi(x)\right)}{\sqrt{2\pi}\alpha_e^{c}
  v_{th,e}^c} e^{-v_x^2/2(v_{th,e}^c)^2}
\Theta\left(1+\frac{v_x}{v_{ec}^{c}}\right) ,\label{fe-cold-dn}
\end{equation}
where $v_{ec}^{c}=\sqrt{2e\Delta\phi/m_e}$ with $\Delta\phi=\phi(x)-\phi_{min}$ and $\alpha_e^{c}=[1+\textup{Erf}(v_{ec}^{c}/\sqrt{2}v_{th,e}^c)]/2$. 
As a result, one has
\begin{align}
  n_{e}^{dn}& = n_{et}^{h}+n_{et}^{c}, \label{eq:ne-coldbeam-Tw-dn} \\
   n_{e}^{dn}V_{ex}^{dn} & =  \frac{n_{et}^{c}v_{th,e}^c}{\sqrt{2\pi}\alpha_e^{c}} e^{-(v_{ec}^{c})^2/2(v_{th,e}^c)^2}-\frac{2n_{et}^{h}v_{th,e}^h}{\sqrt{2\pi}}, \label{eq:nV-coldbeam-Tw-dn}\\
     T_{ex}^{dn} & =  T_c \left[
   \frac{n_{et}^{c}}{n_{e}^{dn}} - \frac{V_{ex}^{dn}v_{ec}^{c}}{(v_{th,e}^c)^2} - \frac{(V_{ex}^{dn})^2}{(v_{th,e}^c)^2}
    \right]\label{eq:neTe-Tw-dn}\\\nonumber
    &+\frac{n_{et}^{h}}{n_e^{dn}}T_h-\frac{2n_{et}^{h}}{\sqrt{2\pi}n_e^{dn}}\frac{v_{ec}^{c}}{v_{th,e}^h}T_h.
\end{align}

Since $V_{ex}^{dn}\ll v_{th,e}^c$, from Eq.~(\ref{eq:nV-coldbeam-Tw-dn}) one finds
\begin{equation}
n_{et}^{c}\approx 2\alpha_e^{c} n_{et}^{h}\sqrt{R}e^{(v_{ec}^{c})^2/2(v_{th,e}^c)^2}= 2\alpha_e^{c} n_{et}^{h}\sqrt{R}e^{e\Delta \phi/T_c},\label{eq:ne-dn-constraint}
\end{equation}
where $R=T_h/T_c$. Near the shock front, $v_{ec}^{c}$ and hence $\Delta \phi$ is approximately zero so that
\begin{equation}
(n_{et}^{c})^{dn}({\rm SF}) \approx (n_{et}^{h})^{dn}({\rm SF})\sqrt{R}\approx 0.2n_h\sqrt{R}= 0.2n_c/\sqrt{R},\label{eq:ne-SF-dn}
\end{equation}
and 
\begin{equation}
T_{ex}^{dn}({\rm SF})  \approx \frac{(n_{et}^{h})^{dn}({\rm SF})}{n_e^{dn}}T_h\approx \sqrt{T_cT_h},\label{eq:Te-SF-dn}
\end{equation}
agreeing well with the simulations.  Notice that the last equality in Eqs.~(\ref{eq:ne-SF-dn})
is only valid for the balanced-pressure case. Compared to the upstream
$T_{ex}^{up}({\rm SF}) \sim 0.5T_h$ in Eq.~(\ref{eq:Tepara-up}),
Eq.~(\ref{eq:Te-SF-dn}) shows that there is a deep cooling of
electrons across the shock front~\cite{zhang23cooling}. More
importantly, Eqs.~(\ref{eq:ne-SF-dn},~\ref{eq:Te-SF-dn}) indicate that
the shock structures have a weak dependence on the initial cold plasma
density and hence the pressure, but are mainly determined by the cold
plasma temperature for a given hot plasma. The requirement for the
initial cold plasma density is simply $n_c\gg n_{e}^{dn}({\rm
  SF})\approx 0.2n_h\sqrt{R}$, i.e., $n_c/n_h\gg 0.2 \sqrt{T_h/T_c}.$

From Eqs.~(\ref{eq:ne-SF-dn},~\ref{eq:Te-SF-dn}), one obtains
\begin{equation}
p_e^{dn}({\rm SF}) =n_e^{dn}({\rm SF})T_{ex}^{dn}({\rm SF}) \approx 0.2n_hT_h= 0.2n_cT_c,\label{eq:pe-dn}
\end{equation}
which agrees well with the simulations. Again, the equality is for the
balanced-pressure case. As a result, in the balanced-pressure case
$p_0=n_cT_c=n_hT_h$, the pressure drop in the downstream from the
downstream recession front $p_e\approx p_0$ to the near shock front
$p_e^{dn}({\rm SF}) \approx 0.2p_0$ is similar to that in the upstream
as indicated in Eq.~(\ref{eq:Tepara-up}), where $p_e^{up}({\rm SF})
=n_e^{up}({\rm SF})T_{ex}^{up}({\rm SF})\approx 0.25p_0$. However, the
corresponding electrostatic potential drops are quite different $-
e\Delta\phi^{dn}\sim T_c\ll T_h\sim e\Delta\phi^{up}$ as shown in
Fig.~\ref{fig:fvx-x}. The upstream potential drop can be easily
understood as a result of the tail hot-electron loss to the cold
plasma as in the sheath case~\cite{tang2016kinetic}. Whereas, the
downstream one can be obtained from Eq.~(\ref{eq:ne-dn-constraint}) in
the limit of $n_{et}^{c}\approx n_c$ near the downstream recession
front
\begin{equation}
- \frac{e\Delta \phi^{dn}}{T_c}\approx \ln\left(\frac{n_c}{2n_{et}^{h}}\frac{1}{\sqrt{R}}\right)\approx 0.9+ 0.5\ln\left(R\right),
\end{equation}
which has a logarithmic and hence weak dependence on the temperature
ratio $R$. This can explain the weak dependence of $U_{SF}$ on $R$ in
Fig.~\ref{fig:cooling-front-width}.

Under the conditions of $V_{ex}^{dn}\ll v_{th,e}^c$ and
$v_{ec}^{c}\lesssim v_{th,e}^c$, we can also obtain
\begin{align}
   q_{en}^{dn}&\approx -\frac{4n_{et}^{h}v_{th,e}^hT_h}{\sqrt{2\pi}}+ \frac{2n_{et}^hv_{th,e}^hT_c}{\sqrt{2\pi}}\left(\frac{(v_{ec}^{c})^2}{(v_{th,e}^c)^2}+2\right)\label{eq:qen-dn}\\\nonumber&\left[\left(\frac{(v_{ec}^{c})^2}{(v_{th,e}^c)^2}+2\right)\frac{T_c}{T_{ex}^{dn}} -3\right]
  n_e^{dn}V_{ex}^{dn}T_{ex}^{dn}.
\end{align}
Similar to the upstream case, the first term on the right-hand-side
(RHS) dominates $q_{en}^{dn}$, $q_{en}^{dn}\sim
-4\times0.2n_hv_{th,e}^hT_h/\sqrt{2\pi}\approx
-0.3p_0v_{th,e}^h$. However the spatial gradient of
$n_{et}^hv_{th,e}^h$ is set by $n_e^{dn}V_{ex}^{dn}$ as indicated by
Eq.~(\ref{eq:nV-coldbeam-Tw-dn}) so that the spatial gradient of
$q_{en}^{dn}$ follows the convective scaling as well.

\subsection{Ion dynamics}
We can perform a similar analysis for the ions, assuming the
distribution as (e.g., see Fig.~\ref{fig:fvx-x})
\begin{align}
f_i &= n_{ib}^c\delta(v_x-v_{ib}^c) \label{eq:truncated-fi-Tw}\\\nonumber
&+\frac{n_{iM}^h}{\sqrt{2\pi}
  v_{th,i}^h \alpha_{i}^h} e^{-(v_x+U_{is}^h)^2/2(v_{th,i}^h)^2}
\Theta\left(-1-\frac{v_x+U_{is}^h}{v_{ic}^h}\right).
\end{align}
Here the first term on the RHS describes the cold ion component in the
downstream only with $n_{ib}^c$ and $v_{ib}^c$ the beam density and
velocity, respectively. The physics underlying such beam distribution
is that the cold ions undergo the decompressional cooling when being
accelerated by the downstream electrostatic field towards the
shock front. \textcolor{black}{As a result, the initial cold ion
  temperature information is lost since $T_{ix}^c\ll T_c$ (in the
  modeled distribution function of Eq.~(\ref{eq:truncated-fi-Tw}), we take $T_{ix}^c=0$ for simplicity),
  and it thus has a weak contribution to the shock in line with the
  cold-ion case in Fig.~\ref{fig:ne-diff-pressure}.}  The second
shifted cutoff Maxwellian term is for the hot ions that can appear
both downstream and upstream, where $n_{iM}^h$ is the particle
density, $U_{is}^h\sim c_s^h$ is a shifted velocity due to the
ambipolar field acceleration, $v_{ic}^h\sim v_{th,i}^h$ is the cutoff
velocity, and
$\alpha_{i}^h=[1-\textup{Erf}(v_{ic}^h/\sqrt{2}v_{th,i}^h)]/2$. Notice
that such hot ion distribution can actually cover the cold-ion case in
Fig.~\ref{fig:ne-diff-pressure} by taking $v_{th,i}^h\rightarrow
0$. \textcolor{black}{Thus, to make the following analysis clear, we
  will use the nominal hot ion sound speed defined as
  \begin{align}
    c_{s,n}^h\equiv\sqrt{T_h/m_i},
  \end{align}
  for necessary normalizations, which is
  equal to $v_{th,i}^h$ in the standard warm ion cases.}
This is analogous to the nominal cold ion sound speed defined in Eq.~(\ref{eq-cold-ion-sound-nominal}).

The ion density, flow and temperature, by integrating the distribution
in Eq.~(\ref{eq:truncated-fi-Tw}), are
\begin{align}
  n_i = & n_{ib}^c+n_{iM}^h, \label{eq:ni-coldbeam-Tw} \\
  n_iV_{ix} =& n_{ib}^cv_{ib}^c-\frac{n_{iM}^hv_{th,i}^h}{\alpha_{i}^h\sqrt{2\pi}} e^{-(v_{ic}^h)^2/2(v_{th,i}^h)^2}-n_{iM}^hU_{is}^h, \label{eq:nV-coldbeam-Tw-ion}\\
 n_iT_{ix}=&  T_hn_{ib}^cv_{ib}^c\frac{v_{ib}^c+v_{ic}^h+2U_{is}^h}{(v_{th,i}^h)^2}-T_hn_{i}V_{ix}\frac{V_{ix}+v_{ic}^h+2U_{is}^h}{(v_{th,i}^h)^2}\label{eq:niTi-coldbeam-Tw} \\ \nonumber
 &+ n_{iM}^hT_h\left[1- \frac{v_{ic}^h+U_{is}^h}{(v_{th,i}^h)^2}U_{is}^h\right].
\end{align}

In the upstream, $V_{ix}^{up}\sim -U_{is}^h$ as shown in
Eq.~(\ref{eq:nV-coldbeam-Tw-ion}) with $n_{ib}^c=0$. Whereas, in the downstream near the shock front,
$V_{ix}^{dn}\approx 0$ so
\begin{equation}
\frac{(v_{ib}^c)^{dn}({\rm SF})}{c_{s,n}^h}\approx \left[\frac{n_{iM}^h}{n_{ib}^c} \left( \frac{v_{th,i}^he^{-(v_{ic}^h)^2/2(v_{th,i}^h)^2}}{c_{s,n}^h\alpha_{i}^h\sqrt{2\pi}} + \frac{U_{is}^h}{c_{s,n}^h}\right)\right]^{dn}(\rm SF).\label{eq:vib-general}
\end{equation}
From the quasi-neutral condition and Eq.~(\ref{eq:ne-SF-dn}), we have $(n_{iM}^h)^{dn}({\rm SF})\sim 0.5 n_h$ and $(n_{ib}^c)^{dn}({\rm SF})\approx n_{e}^{dn}({\rm SF}) \approx 0.2n_h\sqrt{R}$. Therefore, 
\begin{equation}
    (v_{ib}^c)^{dn}({\rm SF})\sim c_{s,n}^h/\sqrt{R} = c_{s,n}^c, \label{eq:vib-dn}
\end{equation} 
where the sum of the two terms in the bracket on the RHS of
Eq.~(\ref{eq:vib-general}) is of order unity.  Specifically, for
$(U_{is}^h)^{dn}({\rm SF})+(v_{ic}^h)^{dn}({\rm SF})\approx 0.6c_{s,n}^h$ as in Fig.~\ref{fig:fvx-x} for $R=100$, the sum of them is $\sim 1.3$
and thus $v_{ib}^c({\rm SF})\sim 3c_{s,n}^c$, agreeing with the
simulation result.

In contrast to the electron case, the ion temperature across the shock
front can increase from the upstream to the downstream (e.g., see
Fig.~\ref{fig:fvx-x}). This is consistent with the standard shock case
that the upstream ion kinetic energy will turn into the downstream
plasma thermal energy across the shock front. Such increase of
$T_{ix}$ can be seen from Eq.~(\ref{eq:niTi-coldbeam-Tw}), where the
second term on the RHS is small in both the upstream and downstream.
Specifically, in the upstream, it is due to the compensation of $V_{ix}^{up}$ with
$v_{ic}^h+2U_{is}^h$; while, in the downstream, $V_{ix}\approx 0$. As
a result, the first term on the RHS of
Eq.~(\ref{eq:niTi-coldbeam-Tw}), which only appears in the downstream,
lead to $T_{ix}^{dn}>T_{ix}^{up}.$

The convective scaling of the ion thermal conduction $q_{in}$ is
trivial. This is because all the characteristic velocities, 
including the flow velocities, in $q_{in}$
\begin{align}
     q_{in} =  & -\frac{n_{iM}^h v_{th,i}^h T_h}{\sqrt{2\pi}\alpha_i^h}\left[\frac{(v_{ic}^h)^2+3(U_{is}^h)^2+3U_{is}^hv_{ic}^h}{(v_{th,i}^h)^2}+2\right] e^{\frac{-(v_{ic}^h)^2}{2(v_{th,i}^h)^2}}\\\nonumber
     &+n_{ib}^cT_c\frac{(v_{ib}^c)^3}{(v_{th,i}^c)^2}-n_{iM}^hT_hU_{is}^h\left[\frac{(U_{is}^h)^2}{(v_{th,i}^h)^2}+3\right]\\\nonumber &-3n_iT_{ix}V_{ix} + 2n_iT_h \frac{V_{ix}^3}{(v_{th,i}^h)^2},
\end{align}
is proportional to $1/\sqrt{m_i}$.

\subsection{Shock front speed}

The shock front speed can be estimated by considering the cold ion continuity and momentum equations, where, in contrast to the distribution function in Eq.~(\ref{eq:truncated-fi-Tw}), we will retain the cold ion temperature $T_{ix}^c$ and discuss its impact,
\begin{align}
\label{eq-ion-density-shock-nc}
& \frac{\partial }{\partial t}n_{ib}^c + \frac{\partial}{\partial x} \left(n_{ib}^c v_{ib}^c\right) = 0, \\
  & m_{i} n_{ib}^c\left(\frac{\partial }{\partial t} v_{ib}^c+ v_{ib}^c\frac{\partial}{\partial x} v_{ib}^c \right)-en_{ib}^cE_x + \frac{\partial }{\partial x}(n_{ib}T_{ix}^c)= 0.\label{eq-ion-momentum-shock-nc}
\end{align} 
 The electric field is provided by the electron force balance of $-en_e^{dn}E_x-\partial p_{e}^{dn}/\partial x\approx 0$ with $p_{e}^{dn}=n_e^{dn}T_{ex}^{dn}$, where $n_e^{dn}\approx n_{ib}^c\gg n_{iM}^h$. To close the equation, a closure for $T_{ex}^{dn}$ is needed. Considering the convective scaling of $dq_{en}^{dn}/dx\propto V_{ex}^{dn}$, we can assume the same length scale for $T_{ex}^{dn}$ and $n_e^{dn}$ so that
\begin{equation}
    \frac{d\ln T_{ex}^{dn}}{dx} = \sigma_e  \frac{d\ln n_{e}^{dn}}{dx}.\label{eq:dTe-dx-dne-dx-dn}
\end{equation}
Similar scale length can be assumed for $T_{ix}^c$ where
\begin{equation}
    \frac{d\ln T_{ix}^{c}}{dx} = \sigma_i^c  \frac{d\ln n_{ib}^{c}}{dx}.\label{eq:dTic-dx-dn}
\end{equation}
Then if one seeks self-similar solutions of $n_{ib}^c$, $v_{ib}^c$, $T_{ix}^c$, and $T_{ex}^{dn}$ as functions of the self-similar variable~\cite{zhang23cooling}, $\xi=x/t$, it can shown that
\begin{equation}
\xi=v_{ib}^c-C_{sb},\label{eq:U-V-cold-ion}
\end{equation}
\textcolor{black}{where
\begin{align}
  C_{sb}=\sqrt{[(1+\sigma_e)T_{ex}^{dn}+(1+\sigma_i^c)T_{ix}^c]/m_i} \label{eq:Csb}
\end{align}
  denotes the cold ion sound speed that takes into account the
  transport effect through $\sigma_e$ and $\sigma_i^c$ (the isothermal
  plasma case corresponds to $\sigma_e=\sigma_i^c=0$).} In fact, $\xi$
is the expansion or recession speed of the plasma so that the shock
front speed reads
\begin{equation}
U_{SF}=\xi({\rm SF})=(v_{ib}^c)^{dn}({\rm SF})-(C_{sb})^{dn}({\rm SF}),\label{eq:Usf-cold}
\end{equation}
where $(v_{ib}^c)^{dn}({\rm SF})\propto c_{s,n}^c$ as shown in Eq.~(\ref{eq:vib-dn}).

To obtain $\sigma_e^{dn}(\rm SF)$ and hence $\left(C_{sb}\right)^{dn}(\rm SF)$, we consider $n_e^{dn}T_{ex}^{dn}$ from Eq.~(\ref{eq:neTe-Tw-dn}), where, near the shock front,
\begin{align}
n_e^{dn}T_{ex}^{dn}&\approx n_{et}^cT_c+ n_{et}^hT_h\approx n_{et}^hT_h,\\
\frac{d \ln(n_e^{dn}T_{ex}^{dn})}{dx}&\approx \frac{n_{et}^cT_c}{n_{et}^hT_h}\left(\frac{d\ln n_{et}^c}{dx}+\frac{n_{et}^hT_h}{n_{et}^cT_c}\frac{d \ln n_{et}^h}{dx}\right).\label{eq:dlnneTe}
\end{align}
The second term in the bracket on the RHS of Eq.~(\ref{eq:dlnneTe}) can be estimated through the force balance of the hot electrons in the downstream
\begin{equation}
   \frac{\partial p_{et}^{h}}{\partial x}\sim -en_{et}^hE_x\sim \frac{n_{et}^h}{n_{e}^{dn}}\frac{\partial p_{e}^{dn}}{\partial x}\sim(1+\sigma_e)n_{et}^hT_{ex}^{dn}\frac{\partial \ln n_{e}^{dn}}{\partial x},
\end{equation}
and thus
\begin{equation}
   \frac{\partial \ln n_{et}^{h}}{\partial x}\sim(1+\sigma_e)\frac{n_{et}^hT_{ex}^{dn}}{n_{et}^{h}T_{ex}^h}\frac{\partial \ln n_{e}^{dn}}{\partial x}\sim (1+\sigma_e)\sqrt{\frac{T_c}{T_h}}\frac{\partial \ln n_{e}^{dn}}{\partial x},\label{eq:partial-neth}
\end{equation}
where the temperature of the hot electrons is $T_{ex}^h\sim T_h.$ Therefore, the second term in the bracket on the RHS of Eq.~(\ref{eq:dlnneTe}) will be the same order or smaller than the first term since $n_{et}^hT_h/(n_{et}^cT_c)\sim \sqrt{T_h/T_c}$ and $n_{et}^c\approx n_e^{dn}$. Thus, from Eq.~(\ref{eq:dlnneTe}) we have
\begin{equation}
\frac{d \ln(n_e^{dn}T_{ex}^{dn}))}{dx}\approx \sqrt{\frac{T_c}{T_h}}\frac{d\ln n_e^{dn}}{dx} \rightarrow (1+\sigma_e)|_{\rm SF} \approx \sqrt{\frac{T_c}{T_h}},
\end{equation}
which implies $\sigma_e \rightarrow -1.$
This suggests that $T_h\partial \ln n_{et}^h/\partial x\sim T_c \partial \ln n_e^{dn}/\partial x $ as seen from Eq.~(\ref{eq:partial-neth}) so that the second term in the bracket on the RHS Eq.~(\ref{eq:dlnneTe}) can be negligible.

More importantly, \textcolor{black}{since $T_{ex}^{dn}(\rm {SF})\approx \sqrt{T_cT_h} $ as shown in Eq.~(\ref{eq:Te-SF-dn}), the contribution of cold ions to the cold ion sound speed is negligible near the shock front as $T_{ix}^c\ll T_c$.} As a result, $(C_{sb})^{dn}({\rm SF})\sim c_{s,n}^c$  and so is the shock front speed
\begin{equation}
    U_{SF}\sim c_{s,n}^c.
\end{equation}
as seen from Eq.~(\ref{eq:Usf-cold}).

\section{Conclusions \label{sec:conclusion}}
In conclusion, the physics of parallel electrostatic collisionless
shocks in hot-cold ablative mixing plasmas has been investigated using
1D kinetic VPIC simulations, revealing a distinct class of
collisionless shocks that fundamentally differs from conventional
cases of over-pressured plasma expansions. One of the key findings is
the weak dependence of such shock on the downstream plasma pressure,
provided the density ratio between the cold and hot plasmas is
sufficiently large $n_c/n_h\gg \sqrt{T_h/T_c}$. The front width scales
with the upstream Debye length that is determined by the hot plasma
temperature, whereas the shock speed is primarily determined by the
cold plasma temperature.

The simulations also highlight the complex electron dynamics across the shock front. Particularly, in the plasma density varying layer between the upstream and downstream recession fronts, the collisionless electron thermal conduction flux follows the free-streaming limit itself but exhibits convective scaling in its spatial gradient. This behavior ensures consistent length scales for electron temperature and density evolution. The non-monotonic electrostatic potential structure, a dynamic double-layer configuration with $e\Delta\phi^{up}\sim T_h$ and $e\Delta\phi^{dn}\sim T_c$, further underscore the unique characteristics of such shock: both upstream and downstream ions are accelerated towards the shock front, while both electrons are trapped. Across the shock front, the electron temperature has a deep drop from the upstream to the downstream, while the ion temperature has an increase due to the conversion of upstream plasma kinetic energy into the thermal energy via the collisionless mixing. 

It should be noted that the complex plasma distributions can destabilize the plasma that may affect the shock dynamics. For the electron modes, this can include the whistler instability in the upstream as in the thermal quench case~\cite{zhang2024collisionless} and the electron beam instabilities. But more importantly, the mixing ions in the downstream can result in ion acoustic instability that leads to an intermittent emission of cold ion beams into the upstream from the shock front, which will be discussed in a separated paper. Notice that these instabilities may be stabilized by even weak collisions~\cite{auerbach1977collisional,zhang2023collisional} so that the consideration of collisions is critical. The effects of collisions on the shock formation are also interesting, which may smear the shock front and affect its speed.

\textbf{Acknowledgment} We thank the U.S. Department of Energy Office
of Fusion Energy Sciences through the Base Fusion Theory Program at
Los Alamos National Laboratory (LANL) under contract
No. 89233218CNA000001. This research used resources of the National
Energy Research Scientific Computing Center, a DOE Office of Science
User Facility supported by the Office of Science of the
U.S. Department of Energy under Contract No. DE-AC02-05CH11231 using
NERSC award FES-ERCAP0032298 and LANL Institutional Computing Program,
which is supported by the U.S. Department of Energy National Nuclear
Security Administration under Contract No. 89233218CNA000001.

\bibliography{reference}

\end{document}